\documentclass[runningheads,a4paper]{llncs}

\def\todoFlag{OFF} 
\def\longFlag{OFF} 
\def\nuevoFlag{OFF} 
\def\borradoFlag{OFF} 

\usepackage{amsmath}

\usepackage{proof}
\usepackage{amsfonts}
\usepackage{latexsym}
\usepackage{amssymb}
\usepackage{amsmath}
\usepackage{alltt}
\usepackage{stmaryrd}
\usepackage{color}

\usepackage{ifthen} 

\newcommand{\nc}{\newcommand}
\nc{\coment}[1]{}

\nc{\bcen}{\begin{center}}
\nc{\ecen}{\end{center}}

\nc{\tp}[1]{\overline{#1}}

\nc{\ra}{\rightarrow}
\nc{\crwlto}{\rightarrowtriangle}  
\nc{\f}{\rightarrow_{\tiny l}}    
\nc{\fcrwl}{\rightarrow_{\it CRWL}}    
\nc{\fc}{\rightarrow_{\it CRWL}}
\nc{\nlet}{\leadsto}        
\nc{\cont}{\hat}

\nc{\gl}{\vdash_{\it CRWL_{\it let}}}
\nc{\bl}{{\cal P} \vdash_{\it BRC_{\it let}}}
\nc{\cl}{{\cal P} \vdash_{\it CRWL}}
\nc{\cld}{{\cal P} \vdash_{\it CRWL'}}
\nc{\fnt}{\rightarrow_{\it let_{nt}}}    
\nc{\fnr}{\leadsto^l}  
\nc{\fnre}{\leadsto^{l^*}}  
\nc{\fnrc}[1]{\leadsto^{l^{#1}}}  
\nc{\jn}{\Join} 
\nc{\tr}{\underline{\mbox{\textbf{t}}}}
\nc{\h}{\rightarrow_{\tiny h}}    
\nc{\dgl}[1]{\den{#1}_{\it GORC_{let}}}
\nc{\ddgl}[1]{\denn{#1}_{\it GORC_{let}}}
\nc{\dg}[1]{\den{#1}_{\it CRWL}}
\nc{\dcl}[1]{\den{#1}_{\it CRWL_{\it let}}}
\nc{\ddcl}[1]{\denn{#1}_{\it CRWL_{\it let}}}
\nc{\tlr}[1]{\widehat{#1}} 
\nc{\rw}{\to} 
\nc{\trw}{\to^r} 
\nc{\tor}{\to} 
\nc{\clto}{\crwlto}                
\nc{\dend}[1]{\den{#1}^d} 
\nc{\cldt}{{\cal P} \vdash_{\it CRWL^{d}}} 
\nc{\con}{{\cal C}}
\nc{\slt}{=} 
\nc{\shlt}{=} 
\nc{\crw}{{\rw^{ct}}}    
\nc{\crwpr}{{\rw_{\cal P}^{{ct}^*}}}    
\nc{\rrw}{{\,\rw^{rt}\,}} 
\nc{\rrwp}{{\,\rw^{rt'}\,}} 
\nc{\rrwpr}{{\rw_{\cal P}^{{rt}^*}}}    
\nc{\Rrw}{{\rw^{Rt}}} 
\nc{\prog}{{\cal P}}
\nc{\crwllet}{{\it CRWL$_{\it let}$\ }}
\nc{\raux}{\rw} 
\nc{\ordap}{\sqsubseteq} 

\newcommand{\var}{\mathcal{V}}

\newcommand{\den}[1]{[\![#1]\!]}
\newcommand{\denn}[1]{[\![\![#1]\!]\!]}

\newcommand{\leqhyp}{\Subset} 

\nc{\sit}[1]{\tau(#1)} 
\nc{\pvar}{PVar} 
\nc{\ohs}{\leqhyp} 
\nc{\hde}{\varphi} 
\nc{\vran}{vran}

\nc{\todo}[1]{\ifthenelse{\equal{\todoFlag}{ON}}{~\\\textcolor[rgb]{1.00,0.00,0.00}{{\fbox{\begin{minipage}{.985\textwidth}{#1}\end{minipage}}}}}{}
} 
\nc{\longu}[1]{\longd{#1}{}} 
\nc{\longd}[2]{\ifthenelse{\equal{\longFlag}{ON}}{{#1}}{{#2}}} 
\nc{\nuevo}[1]{\ifthenelse{\equal{\nuevoFlag}{ON}}{{\textcolor[rgb]{0.00,0.00,1.00}{#1}}}{{#1}}} 
\nc{\borrado}[1]{\ifthenelse{\equal{\borradoFlag}{ON}}{{\textcolor[rgb]{0.95,0.44,0.51}{\emph{[#1]}}}}{}} 
\nc{\nuevoo}[1]{{#1}}
\nc{\borradoo}[1]{}

\nc{\pegado}[1]{\vspace{-0.15cm}\begin{center}#1\end{center}\vspace{-0.2cm}}
\nc{\pop}{\rightarrowtail} 
\nc{\rtcterm}{RtCTerm} 
\nc{\rtexpr}{RtExpr} 
\nc{\rtcsubst}{RtCSubst} 
\nc{\rtm}{{rt}} 
\nc{\rrt}{{rRt}} 
\nc{\rrtHat}{{\_}} %
\nc{\rrtT}[1]{\rrt T({#1})}

\title{A Lightweight Combination of Semantics for Non-deterministic Functions 
\thanks{This work has been partially supported by the Spanish projects
MERIT-FORMS-UCM (TIN2005-09207-C03-03), FAST-STAMP (TIN2008-06622-C03-01/TIN) and Promesas-CAM (S-0505/TIC/0407)
.}
}

\author{Francisco Javier L\'opez-Fraguas \and Juan
  Rodr\'{\i}guez-Hortal\'a \and\\ Jaime S\'anchez-Hern\'andez}

\authorrunning{F.~J.~L\'opez-Fraguas \and J.~Rodr\'iguez-Hortal\'a
  \and J.~S\'anchez-Hern\'andez}

\institute{Departamento de Sistemas Inform\'aticos y
  Computaci\'on\\Universidad Complutense de Madrid, Spain\\
  \email{fraguas@sip.ucm.es, jrodrigu@fdi.ucm.es, jaime@sip.ucm.es}}

\begin{document}

\maketitle
\setcounter{page}{52}

\begin{abstract}
The use of non-deterministic functions is a distinctive feature of modern
functional logic languages. The semantics commonly adopted is {\em  call-time choice},
a notion that at the operational level is related to the {\em  sharing} mechanism of lazy
evaluation in functional languages. However, there are situations where {\em run-time choice},
closer to ordinary rewriting,
is more appropriate. In this paper we propose an extension of  existing call-time choice based languages,
to provide support for run-time choice in localized parts of a program.
The extension is remarkably  simple at three relevant levels: syntax, formal operational calculi
and implementation, which is based on the system \emph{Toy}.
\end{abstract}


\section{Introduction}\label{intro}
\nc{\code}[1]{{\textit{#1}}}
\nc{\scode}[1]{{\small\textit{#1}}}

{Non-strict non-deterministic functions are a distinctive feature of modern functional logic languages
(see \cite{Hanus07ICLP} for a recent survey). It is known that the introduction of non-determinism in a functional setting gives rise
to a variety of semantic decisions (see e.g. \cite{Sondergaard95})}.
For term-rewriting based specifications,
Hussmann \cite{hussmann93} established a major distinction between \emph{call-time choice} and \emph{run-time choice}.
Call-time choice is closely related to call-by-value and, in the case of strict semantics,
it is easily implemented by innermost rewriting. In the case of
non-strict semantics, things are more complicated, since the call-by-value view of call-time choice must
include partial values. Operationally, this
\longu{cannot be achieved by imposing a
particular strategy to ordinary rewriting\footnote{See \cite{ppdp2007} for a
  simple proof of the fact that ordinary rewriting cannot express call-time choice in exact terms.}
and}
needs something similar to the sharing mechanism followed,
for efficiency reasons, in (deterministic) functional languages under lazy evaluation.
In contrast, run-time choice does not share, corresponds rather to call-by-name, and is realized by ordinary rewriting.
For deterministic programs, run-time and call-time are able to produce the same set of values,
but in general the set of values reachable by run-time choice is larger than that of call-time choice.

Non-deterministic functions with non-strict and call-time choice semantics were
introduced in the functional logic setting with the \emph{CRWL} framework \cite{GHLR99}, in which programs
are possibly non-confluent and non-terminating constructor-based term rewriting systems (\emph{CTRS}). Since then,
they are common part of daily programming in systems like \emph{Curry} \cite{Han06curry} or \emph{Toy} \cite{LS99}.
Run-time choice has been rarely \cite{Antoy97ALP} considered as a valuable global alternative to call-time choice.

However, there might be parts in a program or individual functions for which
run-time choice could be a better option,
and therefore it would be convenient to have both possibilities (run-time/call-time) at programmer's disposal.
{The following example illustrates}
 the interest of combining
both semantics.
\\

\noindent \emph{Example 1.}~~
Modeling grammar rules for string generation can be directly done by {CTRS} like the following (non-confluent and non-terminating) one,
in which we assume that texts (terminals) are represented as strings\longd{, identified with lists of characters}{ (lists of characters)},
that can be concatenated with \longu{the infix operation} \code{++} (defined in a standard way):
\begin{center}
\code{
letter $\ra$ "a" ~....~ letter $\ra$ "z"~~~~
word $\ra$ "\," ~~~ word $\ra$ letter++word
}
\end{center}

\noindent Disregarding syntax, this CTRS is  a  valid program in functional logic systems
like \emph{Curry} or \emph{Toy}. {The program acts as a non-deterministic generator of the texts in the language defined
by the grammar. } Each individual reduction leads to a {string in the language}.

{The generation of palindromes (of even length, for simplicity) could be done by the rewrite rules:}

\begin{center}
\code{
\begin{tabular}{l}
palindrome $\ra$ palAux(word)~~~~~
palAux(X) $\ra$ X ++ reverse(X)\\
\end{tabular}
}
\end{center}
\noindent where \code{reverse} is defined in any standard way.
It is important to remark that the definition of  \code{palindrome/palAux} works fine only if call-time choice is
adopted for non-determinism, meaning operationally that in the (partial) reduction
\begin{center}
  \code{palindrome $\ra$ palAux(word) $\ra$ word ++ reverse(word)}
\end{center}
\noindent
the two occurrences of \code{word} created by the rule of \code{palAux} must be shared.
If run-time choice (i.e., ordinary rewriting) were used,
the two occurrences of \code{word} could follow independent ways, and therefore
\code{palindrome} could be reduced, for instance, to \code{"oops"}, which is not a palindrome.
{Two useful operators to  structure grammar specifications are the alternative `$|$'  and Kleene's `$*$' for repetitions:}
\begin{center}
\code{
X $|$ Y $\ra$ X ~~~
X $|$ Y $\ra$ Y~~~~~~
star X  $\ra$ "\,"~~~
star X  $\ra$ X++star(X)  
}
\end{center}
\noindent
{With them \code{letter} and \code{word} could be redefined as follows:}
\begin{center}
\code{
letter $\ra$ "a" $|$ "b" $|$ ... $|$ "z" ~~~~~~~
word $\ra$ star(letter)
}
\end{center}

The annoying fact is that this does not work! At least not under call-time choice,
which implies that this is an uncorrect definition of \code{star} in systems like \emph{Curry} or \emph{Toy}.
The problem with call-time choice here is that all the occurrences of \code{letter} created
by \code{star} will be shared and therefore \code{word} will only generate words like \emph{aaa},  \emph{nnnn}, \ldots,
made with repetitions of the same letter.
To overcome this problem, we would like that in the definition of \code{word}, the application of the \code{star} operation to the string generator \code{letter} could
follow a run-time choice regime, so that each of the two occurrences of \code{letter} created in the rewriting steps
\begin{center}
  \code{ word $\ra$ star(letter)  $\ra$ letter ++ star(letter) }
\end{center}
could evolve independently.
In our proposed extension this would be expressed by writing the definition of \code{word} as follows:
\begin{center}
  \code{ word $\ra$ star(rt(letter))}
\end{center}
where \code{rt} is a special unary function symbol indicating that its argument (\code{letter} in this case)
is not going to be shared in the evaluation of the surrounding application (\code{star(rt(letter))} in this case).

We remark that in this example neither call-time nor run-time choice are a good
single option as semantics for the whole program. The definition of \code{palindrome} 
requires call-time choice, while the use of \code{star} in \code{word} requires run-time choice.
To the best of our knowledge,
no existing implementation {for  functional logic programming} offers the possibility of combining in
the same program both kind of semantics. This paper addresses that problem at a practical level,
aiming at a solution that can be easily realized by modifying existing Prolog-based functional
logic systems. Although our main interest is easiness of implementation, we provide also
formal calculi attempting to reflect at an abstract level the operational behavior
of the extended language. These  calculi could be the technical basis for a thorough investigation
of the formal properties of our proposal, a matter that is out of the scope of this paper.
\section{A tiny functional logic language with run-time choice annotations}

We shortly present here a functional logic language with run-time choice annotations.
To keep the presentation simple, we consider only a first order untyped core with the usual first order
syntax of term rewriting systems. However, the implementation described in Sect. \ref{implementation}
extends the existing system \emph{Toy}, which is a HO typed system using curried notation.

We consider a signature $\Sigma$ made of constructor symbols $c,d,\ldots \in CS$, function symbols $f,g,\ldots \in FS$,
the special unary symbol $rt$, and a set of variables $X,Y,\ldots \in {\cal V}$. We sometimes write $c\in CS^n$ ($f\in FS^n$) to denote a constructor (function) symbol of arity $n$.
Constructor terms (or c-terms) $t,s,\ldots \in CTerm$ follow the syntax: $t ::= X \mid c(t_1,\ldots,t_n)$, and expressions (with run-time
choice annotations) $e,\ldots \in RtExpr$ follow the syntax: $e ::= X \mid c(e_1,\ldots,e_n) \mid f(e_1,\ldots,e_n) \mid rt(e)$.
An intermediate set between $CTerm$ and $RtExpr$ is the set $RtCTerm$ of annotated c-terms
$RtCTerm \ni t ::= X \mid c(t_1,\ldots,t_n) \mid rt(e)$, where $t_1,\ldots,t_n$ are also from $RtCTerm$ and $e$ is any expression.

A \emph{program} is a set of function defining rules, each of the form
$$f(t_1,\ldots,t_n) \rw e$$
where $(t_1,\ldots,t_n)$ is a linear tuple of c-terms from $CTerm$, and $e$ is any expression from $RtExpr$.
We remark that annotated c-terms play no special role in the syntax of programs, but  play an
important role in the parameter passing mechanism, which informally can be explained as follows:
to  apply a program rule  $f(t_1,\ldots,t_n) \rw e$ to a function application $f(e_1,\ldots,e_n)$,
a matching substitution $\theta$ such that $f(t_1,\ldots,t_n)\theta \equiv f(e_1,\ldots,e_n)$ must exist,
and then $f(e_1,\ldots,e_n)$ reduces to $r\theta$, but following the informal criterion about \emph{sharing}:
the copies of subexpressions $e$ of $f(e_1,\ldots,e_n)$ created in $r\theta$ are not shared --i.e. follow run-time choice--
if $e$ is under a $rt$ annotation, and shared --i.e. follow call-time choice-- otherwise.
These ideas are formalized in the next section in the form of two alternative operational calculi.

\section{Formal operational calculi}\label{calculi}
In this section we will try to design some calculi able to express an extension of the standard call-time choice semantics for FLP \cite{GHLR99}, to support the primitive $\rtm$ for run-time choice evaluation. Our approach to formalize this extension is based in two main ideas:
\begin{itemize}
 \item[$\bullet$] The new calculus will be a modification of the simple rewrite calculus presented in \cite{ppdp2007}. As we will have to express run-time evaluation for parts of the computation, we will need to have partially evaluated expressions at our disposal. A calculus in the line of those used in \cite{GHLR99} would not be a suitable tool, as it returns only partial values for the expressions, but no intermediate states of the computation.
\item[$\bullet$] Instead of giving a semantics for annotations $\rtm(e)$ directly, we will think about it as a syntactic sugar for the annotation of the function symbols that appear in $e$ with a $\rtm$ superscript, indicating that those function symbols will be treated as a constructor symbol as far sharing and parameter passing is concerned. Therefore, an expression containing only variables, constructor symbols and function symbols annotated with $\rtm$ could be copied freely, thus getting a run-time behaviour for it, as a function argument. We write $FS^{\rtm}$ for
    the set of function symbols with superscript $\rtm$, $FS^?$ for $FS \cup FS^{\rtm}$ and $f^?$ for function symbols in $FS^?$, i.e., for
    possibly superscripted function symbols.
\end{itemize}

The desugaring of expressions to eliminate the $\rtm$ primitive transforming it into $\rtm$ annotations is performed according to the following definition:
\begin{definition}[Desugaring of the $\rtm$ primitive]
$$
\begin{array}{lcll}
desugar(rt(X)) & = & X & \mbox{ if } X \in \var\\
desugar(rt(c(e_1, \ldots, e_n))) & = & c(desugar(rt(e_1)), \ldots, desugar(rt(e_n))) & \mbox{ if } c \in CS \\
desugar(rt(f(e_1, \ldots, e_n))) & = & f^{\rtm}(desugar(rt(e_1)), \ldots, desugar(rt(e_n))) & \mbox{ if } f \in FS \\
desugar(rt(rt(e))) & = & desugar(rt(e)) \\
\end{array}
$$
\end{definition}

According to this syntactic desugaring for $rt(e)$, the syntax of annotated c-terms and expressions can be reformulated as follows:
\begin{itemize}
 \item[$\bullet$] $\rtcterm \ni t ::= X~|~c(t_1, \ldots, t_n)~|~f^\rtm(t_1, \ldots, t_n)$, if $X \in \var$, $c \in CS^n$, $f \in FS^n$, $t_1, \ldots, t_n \in \rtcterm$
 \item[$\bullet$] $\rtexpr \ni e ::= X~|~c(e_1, \ldots, e_n)~|~f^?(e_1, \ldots, e_n)$, if $X \in \var$, $c \in CS^n$, $f^? \in FS^?$, $e_1, \ldots, e_n \in \rtexpr$
\end{itemize}

To express parameter passing in function applications with $\rtm-$annotated arguments we will need to consider
$rt$-c-substitutions, defined  by: $\theta \in \rtcsubst$ iff $X\theta \in \rtcterm, \forall X\in {\cal V}$.

Now we will define  calculi to work with  annotated expressions. In \cite{ppdp2007} two rewrite notions for call-time choice were defined, each of them being interesting for different applications.  Here we will modify both of them to get two (hopefully) equivalent characterizations of a semantics for annotated run-time choice under a call-time choice environment.
\begin{figure*}[thb]
\framebox{
\begin{minipage}{0.98\textwidth}
\begin{center}
\begin{tabular}{l@{~~}l@{~~~}l}
     \textbf{(B)} & ${\cal C}[e] \pop {\cal C}[\perp]$ & for any context ${\cal C}$ and expression $e \in \rtexpr_\perp$\\[0.15cm]
     \textbf{(OR)} & ${\cal C}[f^?(\overline{p})\theta] \pop {\cal C}[r\theta]$ & for any context ${\cal C}$, $(f(\overline{p}) \tor r \in) \prog$, and $\theta \in \rtcsubst_\perp$
\end{tabular}
\end{center}
\end{minipage}
}
\caption{A one-step reduction relation for non-strict call-time choice with $\rtm$ annotations}\label{br}
\label{popeye}
\end{figure*}

The first characterization is shown in Fig. \ref{popeye}. Its drawback is that the rule \textbf{(B)} involves a `magical' guessing in advance of the fact
that the reduction of a (sub)-expression is not going to be
needed, and replaces this `no need of reduction in the future' by an artificial anticipated
reduction to the undefined value $\perp$.
However, because of its simplicity, the relation is helpful to understand what are the possible results of a reduction.

The second characterization is the
rewrite relation of Fig. \ref{letrcalc}. It expresses in a more realistic manner (specially, if a reduction strategy would be added, which is not
our focus here) the way in which computations are to be performed. To express sharing (when needed), local bindings are created
via a \emph{let} construct.

\begin{figure*}[thb]
\framebox{
\begin{minipage}{0.98\textwidth}
\begin{center}
  \begin{description}
     \item[(Fapp)] $f^?(\overline{p})\theta~\f~r\theta$, ~~~ if $(f(\overline{p}) \tor r) \in \prog$, $\theta \in \rtcsubst$\\[0.15cm]
     \item[(LetIn)] $h(\ldots, e, \ldots) \f let~X=e~in~h(\ldots, X, \ldots)$, if $h\in \Sigma$, $e\equiv f(\overline{e'})$ with $f\in FS$ or $e\equiv let~Y=e'~in~e''$, and $X$ is a fresh variable\\[0.15cm]
     \item[(Bind)] $let~X=t~in~e~\f~ e[X/t]$, ~~~ if $t \in \rtcterm$\\[0.15cm]
     \item[(Elim)] $let~ X = e_1~in~e_2 \f e_2$, ~~~ if  $X \not\in FV(e_2)$\\[0.15cm]
     \item [(Flat)] $let~X = (let~Y =e_1~in~e_2)~in~e_3 ~\f~ let~Y = e_1~in~(let~X=e_2~in~e_3)$\\
 if $Y \not\in FV(e_3)$ \\[0.15cm]
     \item[(Contx)] ${\cal C}[e] \f {\cal C}[e']$, ~if ${\cal C} \neq [\ ]$, $e  \f e'$ using any of the previous
     rules, and in case $e \f e'$ is a (Fapp) step using $(f(\overline{p}) \tor r)\theta \in [{\cal P}]$ then $vRan(\theta|_{\setminus var(\overline{p})}) \cap BV({\cal C}) = \emptyset$.
\end{description}
\end{center}
\end{minipage}
}
\caption{Rules of {\em let}-rewriting extended with $\rtm$ annotations}\label{letRwRt}
\label{letrcalc}
\end{figure*}

Note how in the rule (LetIn), in the case a function application is extracted to a {\it let}, it is needed that $f$ is not marked with $\rtm$, which tell us that it is not allowed to duplicate it, and therefore it may be needed to put it in a \emph{let} in order to progress with the evaluation (for example if it appears in an argument of another function application whose reduction is needed).
\begin{example}\label{Ex1}
Given the program
$$
\begin{array}{ll}
coin \tor 0 ~~~~~& f(X) \tor g(X, coin)\\
coin \tor 1 ~~~~~& g(X,Y) \tor (X,X,Y,Y)
\end{array}
$$
we want to evaluate the expression $rt(f(coin))$, which is desugared as $f^\rtm(coin^\rtm)$. With the calculus of Fig. \ref{popeye} we can do:
$$
\begin{array}{l}
f^\rtm(coin^\rtm) \pop g(coin^\rtm, coin) \pop g(coin^\rtm, 0) \pop (coin^\rtm, coin^\rtm, 0, 0) \\
\pop (0, coin^\rtm, 0, 0) \pop (0, 1, 0, 0)
\end{array}
$$
Note how in the first step the expression $f^\rtm(coin^\rtm)$ can be evaluated as every function symbol present in $coin^\rtm$ is annotated with $\rtm$. On the other hand we cannot apply (OR) to $g(coin^\rtm, coin)$, as one of its arguments contains a function symbol that it is not annoted for run-time, and thus the value $(0, 1, 0, 1)$ is not reachable from $f^\rtm(coin^\rtm)$. This is even more evident in the version of this evaluation got with the calculus of Fig. \ref{letRwRt}:
$$
\begin{array}{l}
f^\rtm(coin^\rtm) \f g(coin^\rtm, coin) \f let~X=coin~in~g(coin^\rtm, X) \\
\f let~X=coin~in~(coin^\rtm, coin^\rtm, X, X) \f let~X=coin~in~(0, coin^\rtm, X, X) \\
\f let~X=coin~in~(0, 1, X, X) \f let~X=0~in~(0, 1, X, X) \\
\f (0, 1, 0, 0)
\end{array}
$$
When we reach the expression $let~X=coin~in~(coin^\rtm, coin^\rtm, X, X)$ it is clear that the first two components of the tuple may evolve in different ways while the values of the last two components will be shared.
\end{example}

\section{A variant of run-time annotations}\label{variant}
In the present section we will show another primitive to express run-time choice that we will build on top of the previous primitive $\rtm$, through a simple program transformation. We will call that primitive $\rrt$, and define its behaviour by the following inference rule that should be added to the {\it CRWL} logic \cite{GHLR99}:
$$
\infer[~~\mathbf{(\rrt)}]{\prog \vdash_{CRWL} \rrt(e) \clto t}
           {
            e \rw^*_{\cal P'} e'~
          \ ~t \ordap |e'|
           }
$$
where ${\cal P'}$ is the program resulting of adding to ${\cal P}$ the new rule $\rrt(e) \tor e$,
and $e \rw^*_{\cal P'} e'$ indicates that $e'$ can be reached from $e$ by zero or more steps of
ordinary rewriting \cite{BaaderNipkow-98} using the program ${\cal P'}$. The approximation ordering
$~t \ordap t'$ between partial values expresses that $t$ is less defined than $t'$
(see \cite{GHLR99} for details).

The rule $\mathbf{(\rrt)}$ itself is already suggesting a possible implementation for $\rrt$. This implementation will be based on the fact that, for any program in which every function symbol that appears in a right hand side of a program rule is $\rtm$-annotated, the evaluation of an expression that has each of its function symbols $\rtm$-annotated too returns the same results as it was evaluated under run-time choice but discarding the annotations. This ideas are formalized in the following definition:
\begin{definition} Given a {\it CRWL}-program $\prog$:
\begin{itemize}
 \item[$\bullet$] We build the signature of a new program $\rrtHat \prog$ adding to it any constructor symbol in the signature of $\prog$, and for any function symbol $f$ in the signature of $\prog$ considering a fresh function symbol $\rrtHat f$ which we add to the signature of $\rrtHat \prog$.
 \item[$\bullet$] We define the transformation of expressions $\rrt$ as:
$$
\begin{array}{lcll}
\rrtT{X} & = & X & \mbox{ if } X \in \var \\
\rrtT{c(e_1, \ldots, e_n)} & = & c(\rrtT{e_1}, \ldots, \rrtT{e_n}) & \mbox{ if } c \in CS \\
\rrtT{f(e_1, \ldots, e_n)} & = & {\rrtHat f}^{\rtm}(\rrtT{e_1}, \ldots, \rrtT{e_n}) & \mbox{ if } f \in FS \\
\rrtT{\rrtT{e}} & = & \rrtT{e}
\end{array}
$$
  \item[$\bullet$] For any $(f(p_1, \ldots, p_n) \tor r) \in \prog$ we add  the rule $\rrtHat f(p_1, \ldots, p_n) \tor \rrtT{r}$ to $\rrtHat \prog$.
\end{itemize}
Finally, any expression $\rrt(e)$ to be evaluated under $\prog$ is desugared into $\rrtT{e}$ and evaluated under $\prog \uplus \rrtHat \prog$
\end{definition}

\begin{example}
Starting with the program of Example \ref{Ex1} we get the program
$$
\begin{array}{c}
\{coin \tor 0, coin \tor 1, f(X) \tor g(X, coin), g(X,Y) \tor (X,X,Y,Y)\}
\\
\uplus
\\
\{\rrtHat coin \tor 0, \rrtHat coin \tor 1, \rrtHat f(X) \tor {\rrtHat g}^{\rtm}(X, coin), \rrtHat g(X,Y) \tor (X,X,Y,Y)\}
\end{array}
$$ under which we can do:$$
\begin{array}{l}
\rrt(f(coin)) \equiv {\rrtHat f}^{\rtm}(\rrtHat coin^{\rtm}) \pop \rrtHat g^{\rtm}(\rrtHat coin^{\rtm}, \rrtHat coin^{\rtm}) \\
\pop (\rrtHat coin^{\rtm}, \rrtHat coin^{\rtm}, \rrtHat coin^{\rtm}, \rrtHat coin^{\rtm}) \pop^* (0,1,0,1)
\end{array}
$$
\end{example}

\section{Implementation issues}\label{implementation}
In order to study the practicability of the proposal we have implemented it as an
extension of the functional logic system \emph{Toy} (\cite{toyreport}). This system, as well
as
other modern systems like \emph{Curry} (\cite{Han06curry}), operates under call-time
  choice. We
introduce the new syntactic construct {\it rt e} into the syntax of \emph{Toy} to instruct
the system to evaluate the expression {\it e} under a run-time choice
regime. The system will use run-time choice for evaluating the expressions annotated with
{\it rt}, and call-time choice as usual for the rest of computations, i.e., we
have within the same language both regimes of evaluation.

The extension is well supported by the system and requires only some light\-weight
modifications. In fact, the traditional problem is how to achieve sharing
in a non-deterministic language like this, and our goal now is to inhibit
this sharing mechanism at the points required by the programmer with {\it rt}.

\emph{Toy} is implemented in Prolog and uses Prolog as target code (see \cite{LLR93,toyreport} for
details). Sharing is implemented by means of {\it suspensions}, that are Prolog
terms of the form:
\begin{center}
  {\it susp(FunctionName,Arguments,Result,Evaluated)}
\end{center}
The {\it FunctionName} and its {\it Arguments} represent the expression $e$ to
be evaluated, while {\it Result} is the resulting value (if evaluated, variable in
other case) and {\it Evaluated} is a flag that indicates if the expression has
been evaluated (flag {\it on}) or not (flag variable). Every function call is
translated into a suspension in
order to share its value when the expression is passed as argument and copied.
As an example of the use of this representation consider the following program:
\begin{alltt}
  coin = 0
  coin = 1

  double X = X + X

  test1 = double coin
  test2 = rt (double coin)
\end{alltt}
Consider the evaluation of {\it test1}. As all the function calls are translated
into suspended forms, in particular {\it coin} will
have the form {\it susp(coin,[],R,E)}. The evaluation of {\it double} does not
demand the evaluation of its argument {\it coin}, so it will produce
\begin{center}
  {\it susp(coin,[],R,E) + susp(coin,[],R,E)}
\end{center}
Later, when one of the calls to {\it coin} is evaluated, for example to 0, the other
one automatically gets the same value:
\begin{center}
  {\it susp(coin,[],0,on) + susp(coin,[],0,on)}
\end{center}
The result of the addition is 0, that is a value obtained for {\it test1}. If we
evaluate {\it coin} to 1 we have
\begin{center}
  {\it susp(coin,[],1,on) + susp(coin,[],1,on)}
\end{center}
and then result 2, that is the other value obtained for {\it test1}. With this
sharing mechanism we can not
obtain the value 1 for {\it double coin} as it would require to evaluate both calls
to {\it coin} to two different values.

For the function function {\it test2} we would want to obtain the values 0 and 2
as before, but also the value 1 (evaluating separately both calls to {\it
  coin}). In this case {\it rt} will deactivate the sharing
mechanism. This can be easily achieved by translating the call {\it coin} into
the suspended form {\it susp(coin,[],R,rt)}. The flag {\it rt} will indicate to
the system that the
value of this expression must not be shared (and neither kept in the variable
{\it R}). For {\it test2} we evaluate
\begin{center}
  {\it susp(coin,[],R,rt) + susp(coin,[],R,rt)}
\end{center}
The first suspension can be reduced to 0 (without annotating the result in {\it
  R}), and the second one to 1, obtaining 1 for {\it test2} as expected.

The extension implemented in \emph{Toy} provides this behaviour with {\it test1} and
{\it test2}. In fact, for {\it test2} it obtains 0, 2 and 1 twice (evaluating
the first {\it coin} to 0 and the second to 1 and viceversa). As another
example, consider the problem of generating numbers as combinations of the
digits 0, 1 and 2. Using {\it take}, {\it repeat} and the alternative
operator `$|$' (introduced in Sec. \ref{intro}) we could define:
\begin{alltt}
  number N = take N (repeat (0 | 1 | 2))
\end{alltt}
but then the expression {\it number 3} will produce only the answers [0,0,0],
[1,1,1] and [2,2,2], because the expression {\it 0 $|$ 1 $|$ 2} is evaluated only once
and then its value is shared when evaluating {\it repeat}. For achieving the
expected behaviour we have to instruct the system for choosing the digits under
run-time choice (to avoid sharing):
\begin{alltt}
  number N = take N (repeat (rt (0 | 1 | 2)))
\end{alltt}
Now we obtain the 27 possible combinations that include $[1,1,2]$ or $[3,1,2]$
as instance.
The example of palindromes of Sect. \ref{intro} also works as expected.

The prototype and some examples can be found at\\
\texttt{https://gpd.sip.ucm.es/trac/gpd/wiki/GpdSystems}.

\section{Conclusions}\label{conclu}
We have proposed a simple way of combining in the same program run-time choice and
call-time choice, two semantics commonly adopted for non-determinism in
rewriting-based declarative languages, but that cannot coexist within the same
program in current systems.

The approach presented here starts from a call-time choice ambient
(as given by most popular functional logic systems like \emph{Curry} \cite{Han06curry} or \emph{Toy} \cite{LS99})
and adds to it the possibility of annotating the evaluation of (sub)-expressions as following
a run-time choice regime. We have proposed two variants of this idea, the first being more 'local'
in the effect of an annotation $rt(e)$, while the second is more global. In both cases
we have proposed a formal definition of the intended semantics.

For the first variant we have given formal operational descriptions, by adapting to the
new setting two one-step reduction relations proposed in \cite{ppdp2007} as a simple
notion of rewriting adequate for call-time choice. As for implementation, this variant
has been  achieved by modifying of the system \emph{Toy}. Essentially, we have needed to
change the management of \emph{suspensions}, that are the technical key to implement
sharing for call-time choice. The resulting prototype can be found at
\texttt{https://gpd.sip.ucm.es/trac/gpd/wiki/GpdSystems}.

For the second variant we give a logical
semantics that extends, to cope with \emph{rt} annotations, the proof calculus
of the \emph{CRWL} framework \cite{GHLR99}. We have seen how to transform
annotations of this variant into the first one. This mapping can be used to implement
the second variant.

Recently, we have tried a different alternative to the combination
of call-time and run-time choice \cite{LRStchrRTCT08long}, following
a way complementary to the one in this paper: there we start from ordinary rewriting
and enhance it with local bindings through a \emph{let} construct to express sharing
and call-time choice. The resulting framework seems to be more amenable to formal
treatments, as shown by the good number of technical results obtained in
\cite{LRStchrRTCT08long}. On the other hand, the approach here seems to be more
easily implementable, at least if one wants to reuse existing call-time-choice
based implementations.


\end{document}